\documentclass[aps,pre,reprint,showpacs]{revtex4-1}
\usepackage{graphicx}
\usepackage{amsmath}
\usepackage{amssymb}
\usepackage{psfrag}
\usepackage{natbib}

\newcommand{\av}[1]{\langle{#1}\rangle}

\begin{document}
\title{Maximizing coherence of oscillations by external locking}
\author{Arkady Pikovsky}
\affiliation{Institute for Physics and Astronomy, 
University of Potsdam, Karl-Liebknecht-Str. 24/25, 14476 Potsdam-Golm, Germany}

\begin{abstract}
We study how the coherence of noisy oscillations can be optimally enhanced 
by external locking. Basing on the condition of minimizing the phase 
diffusion constant, we find the optimal forcing explicitly in the limits of small 
and large noise, in dependence of phase sensitivity of the oscillator. We show that
the form of the optimal force bifurcates with the noise intensity. In the limit
of small noise, the results are compared with purely deterministic conditions of 
optimal locking.
\end{abstract}
\date{\today}
\date{\today}

\maketitle

Autonomous self-sustained oscillations may be extremely regular 
(like, e.g., lasers) or
rather incoherent (like many biological oscillators, e.g., ones responsible for
cardiac or circadian rhythms).
A usual way to improve the quality of oscillations is to lock (synchronize)
them by an external pacing~\cite{Kuramoto-84,Pikovsky-Rosenblum-Kurths-01}.
This is used in radio-controlled clocks and in cardiac pacemakers; also 
circadian rhythms are nearly perfectly locked by the 24-hours day/night force.

In this letter we address a question: which periodic force ensures, via locking,
the maximal coherence
of a noisy self-sustained oscillator? Of course, one has to fix the amplitude of 
the force, so the nontrivial problem is in finding the optimal force profile.
We will treat this problem in the phase approximation~\cite{Kuramoto-84}, 
which is valid for general
oscillators, provided the noise and the forcing are small. In this approximation the
dynamics of the phase reduces to a noisy 
Adler equation~\cite{Pikovsky-Rosenblum-Kurths-01,Goldobin2010}, 
and the maximal coherence
is achieved if the diffusion constant of the phase is minimal. It should be noted
that an optimal locking problem has been recently discussed for purely deterministic
oscillations. There, the optimal condition was formulated as the maximal width of the
Arnold's tongue (the synchronization region) or as the maximal stability of the locked 
state~\cite{Harada_etal-10,Zlotnik_etal-13,Tanaka-14,Hasegawa-Arita-14}. In our 
case there is an additional
parameter, the noise intensity, and we will show that the optimal force profile 
depends on the noise amplitude. Below we will also compare the 
limit of small noise with purely
deterministic setups.

Let us consider a self-sustained oscillator with frequency $\omega$, its phase
in presence of a small Gaussian white noise obeys the Langevin equation
\begin{equation}
\frac{d\varphi}{dt}=\omega+\beta^{-1/2}\xi(t),\qquad 
\langle \xi(t)\xi(t')\rangle=2\delta(t-t'),
\label{eq:p1}
\end{equation}
where $\beta^{-1}$ is the noise intensity. A small periodic forcing 
with frequency $\Omega$ leads, in the first order in the force amplitude, to the
following phase dynamics~\cite{Kuramoto-84,Goldobin2010}:
\begin{equation}
\frac{d\varphi}{dt}=\omega+s(\varphi)f(\Omega t)+\beta^{-1/2}\xi(t).
\label{eq:p2}
\end{equation}
Here $s(\varphi)$ is the phase sensitivity function (a.k.a. phase response curve), 
and $f(\Omega t)$ is 
the phase-projected force term. Our goal will be to find such a forcing
$f(\cdot)$ that 
maximizes the coherence, i.e. minimizes the diffusion constant of the
phase $\varphi$.
This optimal force will depend on the phase sensitivity function $s(\cdot)$ and 
on the noise intensity $\beta$.

As the first step we introduce the slow phase $\phi=\varphi-\Omega t$ and perform 
the standard 
averaging over the period 
$2\pi\Omega^{-1}$~\cite{Kuramoto-84,Pikovsky-Rosenblum-Kurths-01}, this yields
\begin{equation}
\frac{d\phi}{dt}=\omega-\Omega+g(\phi)+\beta^{-1/2}\xi(t)=
-\frac{dv(\phi)}{d\phi}+\beta^{-1/2}\xi(t),
\label{eq:p3}
\end{equation}
where
\begin{equation}
g(\phi)=\frac{1}{2\pi}\int_0^{2\pi}
dy s(\phi+y)f(y),
\label{eq:p4}
\end{equation}
and we introduced the ``potential''
\begin{equation}
v(\phi)=(\Omega-\omega)\phi-\int^\phi g(y)\,dy.
\label{eq:p5}
\end{equation}
Let us consider a situation, where the mean frequency of oscillations is exactly
that of the forcing; this means that the slow phase $\phi$ performs a random walk 
without a bias. This happens for a purely periodic, non-inclined
potential. This condition, as it follows from~\eqref{eq:p5}, 
defines the optimal frequency of the forcing
\begin{equation}
\overline{\Omega}=\omega+\av{s}\av{f},
\label{eq:p51}
\end{equation}
where we denote $\av{f(\phi)}=(2\pi)^{-1}\int_0^{2\pi}f(\phi)\,d\phi$.
Thus, without loss of generality we can assume that $\av{s}=\av{f}=0$
and $\Omega=\omega$.

The problem of finding the diffusion constant $D$ of a particle 
in a periodic potential $v$, driven by a white Gaussian noise, has been 
solved in Ref.~\cite{Lifson-Jackson-62}
(and generalized to the case of an inclined potential in Ref.~\cite{Reimann2001}):
\begin{equation}
D=\frac{D_0}{\av{\exp(\beta v)}\av{\exp(-\beta v)}},
\label{eq:p6}
\end{equation}
where $D_0$ is the bare diffusion constant without potential. Thus, the problem
of maximizing the coherence reduces to maximizing the expression
\begin{equation}
C=\av{\exp(\beta v)}\av{\exp(-\beta v)}.
\label{eq:p7}
\end{equation}
As an additional condition we have to fix the intensity of the force:
\begin{equation}
\av{f^2}=const.
\label{eq:p8}
\end{equation}

The formulated optimization problem is quite complex to be solved in general.
Therefore, below we consider some simplifying cases, and will perform a rather
full analysis for a simple bi-harmonic phase sensitivity function. The main feature 
we will focus on, are bifurcations in dependence on the form of this function and 
on the noise intensity; we will see that different forcing waveforms provide optimal 
coherence in different domains of the parameter space.

For the analytical consideration below it is convenient to use Fourier transforms, 
which we will denote by capitals:
 \begin{equation}
s(x)=\sum_k S_k \exp[ikx],\quad S_k=\frac{1}{2\pi}\int_0^{2\pi} s(x) \exp[-ikx]dx,
\label{eq:p9}
\end{equation}
and the same for functions $f,g,v$, Fourier harmonics of which we denote as
$F_k,G_k,V_k$, 
respectively. Because $g(\phi)$ is according to~\eqref{eq:p4} a convolution of $f$ and $s$,
and $v$ is the integral of $g$, we have
 \begin{equation}
G_k=S_kF_{-k},\qquad V_k=ik^{-1}S_kF_{-k}.
\label{eq:p10}
\end{equation}
The condition on the norm of the force~\eqref{eq:p8} now reads
 \begin{equation}
\sum_k|F_k|^2=const.
\label{eq:p11}
\end{equation}

We start with the case of strong noise (small $\beta$). Expanding~\eqref{eq:p7}, we obtain
a simple expression for the quantity to be maximized:
 \begin{equation}
C\approx 1+\beta^2 \av{v^2}=1+\beta^2\sum_k k^{-2}|S_k|^2 |F_k|^2.
\label{eq:p12}
\end{equation}
Together with condition~\eqref{eq:p11}, the maximum can be found by virtue of Lagrange 
multipliers:
 \begin{equation}
|F_k|\sim\delta_{k,K},\quad\text{where}\quad K=\text{arg max}(k^{-2}|S_k|^2).
\label{eq:p13}
\end{equation}
Thus, for large noise, the optimal forcing is purely harmonic one $f(x)\sim\cos(Kx)$,
where $K$ is determined from~\eqref{eq:p13}.

The case of small noise is the limit $\beta\to\infty$. In this case the integrals
in the expression~\eqref{eq:p7} can be asymptotically estimated as Laplace integrals:
 \begin{equation}
\begin{aligned}
\av{\exp(\beta v)}&\approx (2\pi)^{-1}\exp(\beta v_{max}),\\
\av{\exp(-\beta v)}&\approx (2\pi)^{-1}\exp(-\beta v_{min}),
\end{aligned}
\label{eq:p14}
\end{equation}
what gives
 \begin{equation}
C\sim \exp(\beta (v_{max}-v_{min})).
\label{eq:p15}
\end{equation}
Suppose now that $v_{min}=v(x_2)$ and $v_{max}=v(x_1)$. Then
 \begin{equation}
\begin{gathered}
\ln C\sim v_{max}-v_{min}=\int_{x_1}^{x_2} g(\phi)\,d\phi=\\
=\frac{1}{2\pi}
\int_0^{2\pi} f(x) dx \int_{x+x_1}^{x+x_2}s(y)dy.
\end{gathered}
\label{eq:p16}
\end{equation}
Using additionally condition~\eqref{eq:p8} with a Lagrange multiplier, we obtain
 \begin{equation}
f(x)=\text{const} \int_{x+x_1}^{x+x_2}s(y)dy.
\label{eq:p17}
\end{equation}
 Substituting this into conditions $g(x_1)=g(x_2)=0$, we get an equation for 
$\Delta=x_2-x_1$ (only this difference is important, but not the values of $x_1,x_2$):
\begin{equation}\begin{gathered}
p(\Delta)=\int_0^{2\pi}s(z)\int^{\Delta}_{0}s(z+y)\,dz\, dy=\\
=2\pi\sum_k k^{-1}|S_k|^2\sin k\Delta=0.
\end{gathered}\label{eq:p18}
\end{equation}
This equation has always a solution $\Delta=\pi$, but depending on the
form of the phase sensitivity function $s$
there can be other solutions, corresponding to local maxima of $C$; one has to compare
different possible values of $\Delta$ to find the global maximum. 
Once $\Delta$ is found,
the corresponding force can be expressed as
\begin{equation}
f(\phi)\sim\int_\phi^{\phi+\Delta} s(y)\;dy,\qquad F_k\sim S_k \frac{\exp[ik\Delta]-1}{ik}.
\label{eq:p19}
\end{equation}
 
Here below we present the simplest nontrivial example, where it is possible, in 
addition to the asymptotic cases of small and large noise considered above, 
to perform 
the analysis for intermediate noise levels. We consider the bi-harmonic phase sensitivity 
function
\begin{equation}
s(x)=2\sqrt{q}\cos x+2\sqrt{1-q}\cos 2x,
\label{eq:p20}
\end{equation}
where parameter $q$ describes the relative weight of the harmonics. 

The limit of strong noise~\eqref{eq:p13}, with $|S_1|^2=q$, $|S_2|^2=1-q$, yields
\begin{equation}
f(x)\sim \begin{cases} \cos 2x&\text{ if } 0\leq q<1/5,\\
\cos x&\text{ if } 1/5< q\leq 1.\end{cases}
\label{eq:p23}
\end{equation}

The limit of weak noise leads to the following expression for function~\eqref{eq:p18}:
\begin{equation}
p(\Delta)=q\sin\Delta+\frac{1-q}{2}\sin 2\Delta.
\label{eq:p24}
\end{equation}
For  $q>1/2$, the only root in~\eqref{eq:p24} is $\Delta=\pi$, 
while for $q<1/2$ there is an additional root $\Delta_1=\arccos(-q/(1-q))$.
Substituting this into~\eqref{eq:p19}, we obtain for small noise
\begin{equation}
|F_1|^2=1-|F_2|^2= \begin{cases} 2q&\text{ if } 0\leq q<1/2,\\
1&\text{ if } 1/2< q\leq 1.\end{cases}
\label{eq:p25}
\end{equation}

\begin{figure}[!th]
\centering
\psfrag{b=1}[rr][cc]{$\beta=1$}
\psfrag{beta}[cc][cc]{$\beta$}
\psfrag{ylabel}[cc][cc]{$|F_1|^2$}
\includegraphics[width=0.8\columnwidth]{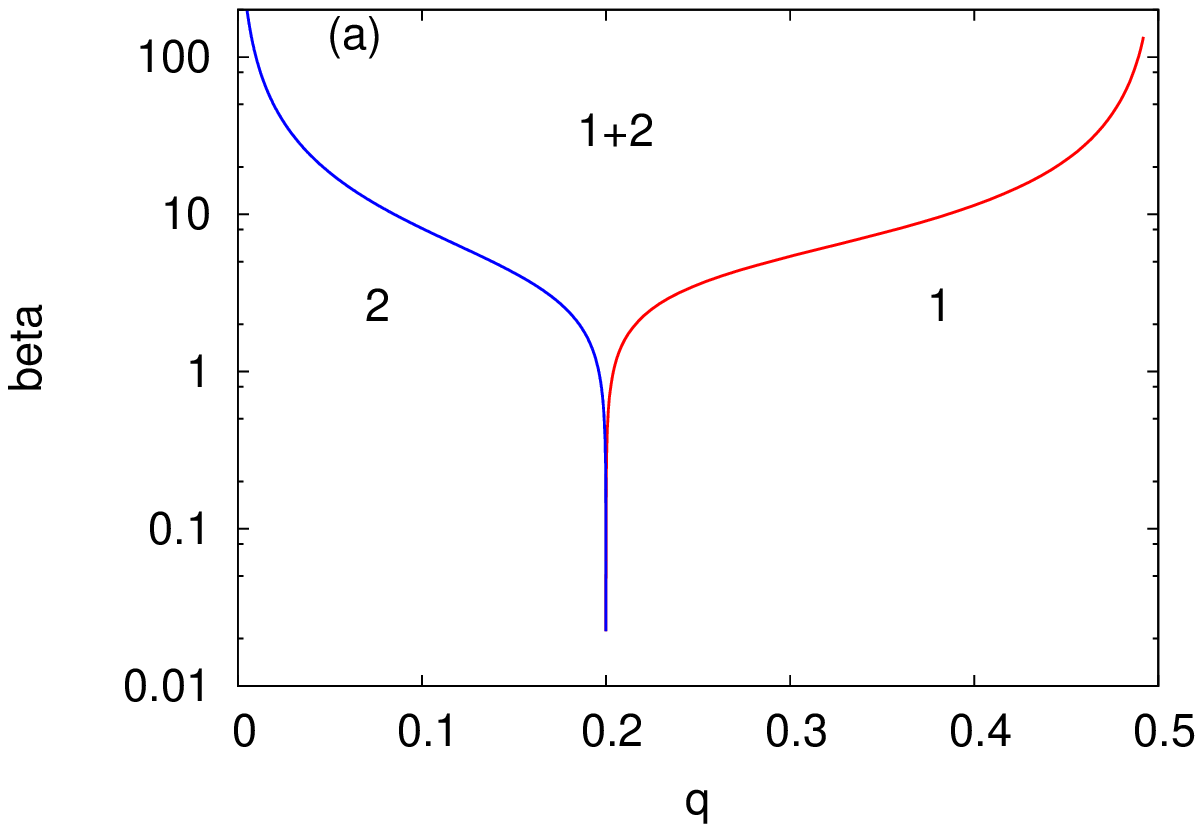}\\
\includegraphics[width=0.8\columnwidth]{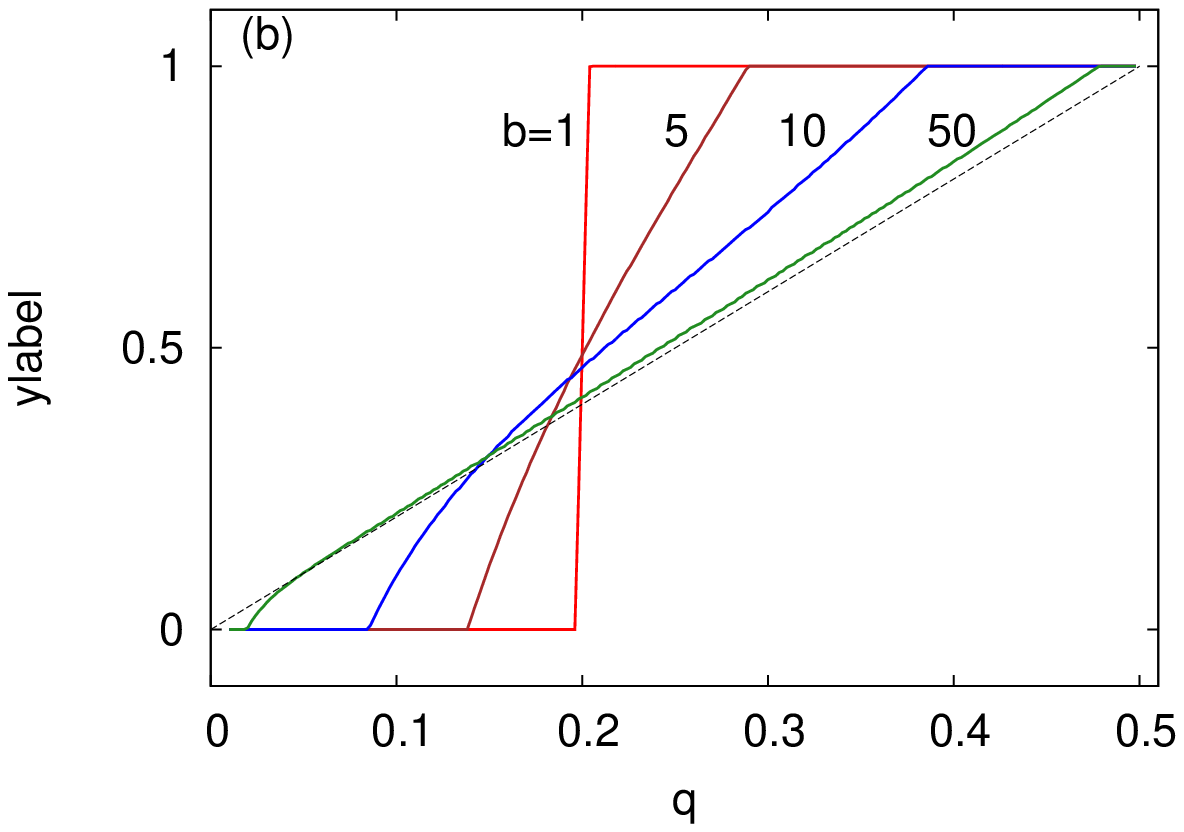}\\
\includegraphics[width=0.8\columnwidth]{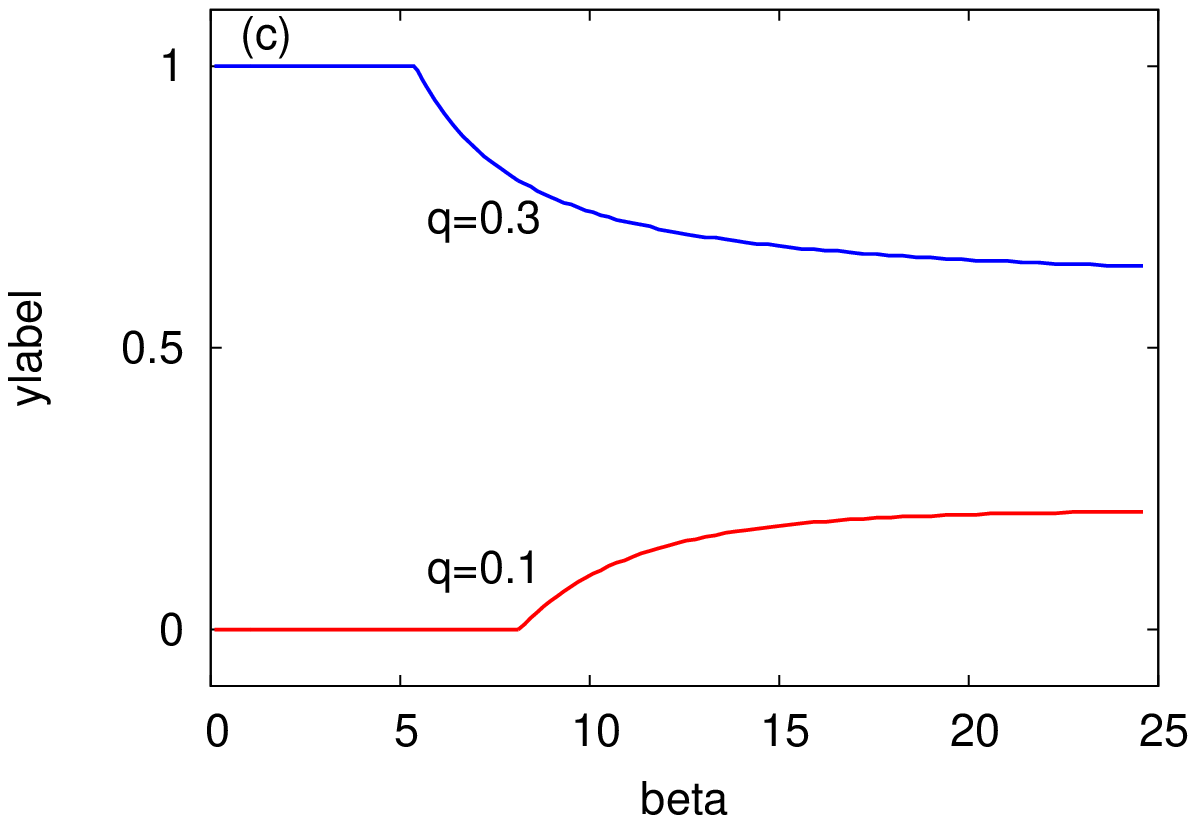}
\caption{(color online) (a) Domains on the plane of parameters $(q,\beta)$ where
the optimal force has one harmonics (2: the second one, 1: the first one), 
and two harmonics (1+2), according top expressions~(\ref{eq:p26},\ref{eq:p27}). 
(b) The intensity of the first harmonics $|F_1|^2$ as a 
function of $q$ for different noise intensities $\beta$. Thin dashed black line shows the 
limit of small noise~\eqref{eq:p25}. (c) Two dependencies of $|F_1|^2$ on the noise 
intensity $\beta$ showing bifurcations from one-mode to two-mode solutions at critical values
of $\beta$.}
\label{fig:1}
\end{figure}

Let us now consider general noise intensities.
The forcing
in this case should be also generally bi-harmonic (higher harmonics disappear 
according to~\eqref{eq:p10}):
\begin{equation}
f(x)=a\cos x+b\cos 2x+c\sin 2x,
\label{eq:p21}
\end{equation}
with unknown constants $a,b,c$ satisfying $a^2+b^2+c^2=1$. In this representation
the potential $v(x)$ reads
\begin{equation}
v(x)=-\sqrt{q}a\sin x-\frac{\sqrt{1-q}b}{2}\sin 2x-\frac{\sqrt{1-q}b}{2}\cos 2x.
\label{eq:p22}
\end{equation}
Unfortunately, after substitution of this potential
in the expression~\eqref{eq:p7} for the factor $C$, we obtain integrals
which cannot be expressed in a closed analytic form. However, for a 
purely first-harmonic 
forcing ($b=c=0$) and a purely second-harmonic forcing ($a=0$), 
the factor $C$ as well as its 
derivatives can be expressed via first order Bessel functions. Thus, it is possible
to find the domains of stability of these pure forcing terms analytically, for 
arbitrary values of noise intensity $\beta$. These lengthy but straightforward 
calculations give the stability boundaries in a parametric form: The first-harmonic force
loses stability at the curve on the $(\beta,q)$ plane given according to
\begin{equation}
q=\frac{z(-I_4(z)+I_0(z))}{8I_1(z)+z(-I_4(z)+I_0(z))}\;,\quad
\beta=\frac{z}{\sqrt{q}}.
\label{eq:p26}
\end{equation}
The stability boundary of the second-harmonic solution is 
\begin{equation}
q=\frac{I_1(z)}{I_1(z)+2zI_0(z)}\,,\quad
\beta=\frac{2z}{\sqrt{1-q}}.
\label{eq:p27}
\end{equation}
We illustrate these domains in Fig.~\ref{fig:1}. Here we also show 
numerically obtained dependencies of  $|F_1|^2$ (the intensity of the second harmonis
is $|F_2|^2=1-|F_1|^2$) on parameters $q$ and $\beta$,
demonstrating bifurcations on the form of the forcing.

\begin{figure}[!tb]
\centering
\psfrag{ylabel}[cc][cc]{$|F_1|^2$}
\includegraphics[width=0.7\columnwidth]{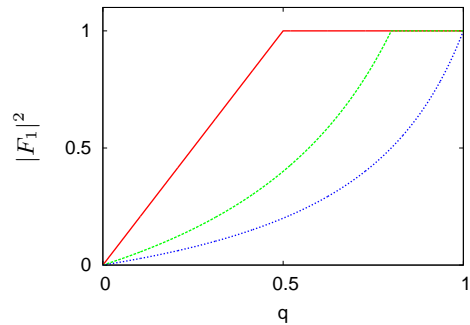}
\caption{(color online) The intensity of the first harmonic component in the optimal force as
function of parameter $q$, for three optimization criteria: top solid red line:
maximal coherence in the weak noise limit~\eqref{eq:p25}; middle dashed green line:
maximal width of the synchronization region~\eqref{eq:p28}; bottom dotted blue
curve: maximal linear stability of the locked state~\eqref{eq:p29}.}
\label{fig:2}
\end{figure}

Next we discuss a relation between different criteria used for the ``optimal locking''.
While here we optimize the coherence in the presence of noise, in 
Refs.~\cite{Harada_etal-10,Zlotnik_etal-13} purely deterministic criteria
have been suggested. It is  instructive to compare them with
our approach in the limit of small noise. Suppose that the coupling
function $g(\phi)$ has zeros at $\phi_{1,2}$ (where $\phi_1$ is the stable one), and
extrema at $\phi_{3,4}$. In the approach of~\cite{Zlotnik_etal-13}, the linear 
stability at the stable equilibrium $|g'(x_1)|$ is maximized. In the approach
of~\cite{Harada_etal-10}, the width of the synchronization region 
$\sim|g(\phi_3)-g(\phi_4)|$ is maximized. In our maximization of the coherence,
the potential barrier for a noise-induced phase slip $\sim\left|\int_{\phi_1}^{\phi_2} g(x)dx\right|$
should be maximal. For the discussed above example of a bi-harmonic phase sensitivity 
function~\eqref{eq:p20}, all the optimal forcings can be found analytically, they are 
generally also bi-harmonic. The approach of~\cite{Zlotnik_etal-13} 
yields in this case 
\begin{equation}
|F_1|^2=1-|F_2|^2=q/(4-3q),
\label{eq:p29}
\end{equation}
while the approach of~\cite{Harada_etal-10} gives
\begin{equation}
|F_1|^2=1-|F_2|^2= \begin{cases} \frac{2q}{4-3q}&\text{ if } 0\leq q<4/5,\\
1&\text{ if } 4/5< q\leq 1.\end{cases}
\label{eq:p28}
\end{equation}
We compare the results in Fig.~\ref{fig:2}. One can see that for the minimal coherence, presence of a strong first
harmonics component in the forcing is more important than for other criteria.

In conclusion, we have studied the problem of maximizing coherence of oscillations 
by external locking, in the phase approximation. The optimal phase
forcing function depends not
only on the phase sensitivity function of the system, but also on the noise intensity.
For large noise a purely harmonic forcing is optimal, the number of the harmonic depends 
on the phase sensitivity. For smaller noise, a bifurcation to a more general, 
multi-harmonic forcing may occur. We have also demonstrated, that different optimality 
conditions in the purely deterministic case lead to different optimal forcing functions,
which also differ from the limit of small noise when optimization of the coherence 
is performed.

The author thanks the Galileo Galilei Institute for
Theoretical Physics, Florence, Italy, for the hospitality and
the INFN for partial support during the completion of this
work.

\end{document}